\documentclass[a4paper,twocolumn]{esapub} 
          [2001/04/25 1.1 (PWD)]

\usepackage{times}
\usepackage{natbib}
\usepackage{graphics}
\usepackage{amssymb} 
\title{Oscillation power spectra of the Sun and of $\alpha$  Cen A: 
observations versus models}
\author{R\'eza Samadi}
\author{M.J. Goupil}
\affil{Observatoire de Paris, 5 place J. Janssen, 92195 Meudon, France}
\author{F. Baudin}
\affil{Institut  d'Astrophysique Spatiale, Orsay, France}
\author{D. Georgobiani}
\affil{Center for Turbulence Research, Stanford University NASA Ames Research Center, Moffett Field, USA}
\author{R. Trampedach}
\affil{ Research School of Astronomy and Astrophysics, Mt. Stromlo Observatory, Weston, Australia}
\author{R. Stein}
\affil{Michigan State University, East Lansing, USA}
\author{A. Nordlund}
\affil{Niels Bohr Institut, Copenhagen, Denmark}

\newcommand{\fig}[3]{
      \begin{figure}[tbp]
	\resizebox{\hsize}{!}{\includegraphics  {#1}}
	\caption{#2}
	\label{#3}
        \end{figure} }
\newcommand{\eqn} [1] {
\begin{equation} 
#1 
\end{equation}}

\begin{document}


\maketitle

\begin{abstract}

Hydrodynamical, 3D simulations of the outer layers of the Sun and $\alpha$\,Cen\,A are used to  obtain constraints on the properties of turbulent convection in such stars. These constraints enable us to compute --- on the base of a theoretical model of stochastic excitation --- the  rate $P$ at which p modes are excited by turbulent  convection in those two stars.
Results are then compared with solar seismic observations and recent observations of $\alpha$\ Cen\ A. For the Sun, a good agreement between observations and computed $P$ is obtained. For   $\alpha$~Cen~A a large discrepancy is obtained which origin cannot be yet identified: it can either be caused by the present data quality which is not sufficient for our purpose or by  the way the intrinsic amplitudes and the life-times of the modes are determined or finally attributed to our present modelling. Nevertheless, data with higher quality or/and  more adapted data reductions will likely provide  constraints on the p-mode excitation mechanism in  $\alpha$~Cen~A.
\end{abstract} 

\section{Introduction}
Solar-like oscillations are stochastically excited by turbulent convection and damped by several mechanisms.
The square of the mode amplitude is proportional to the ratio between the rate $P$ at which the the mode is excited  and the mode damping rate $\eta$.
The latter is proportional to the mode line-width ($\Gamma$) which is  inversely proportional to the mode life-time ($\tau$).

Providing that measurements of the oscillation amplitudes and life-time (or line-width) are available it is possible to compute $P$ and hence to derive constraints on the models of stochastic excitation.
Such measurements have been available for the Sun for several years \citep[e.g. ][]{Libbrecht88,Chaplin97,Chaplin98}.
Recently \citet{Butler04} have obtained ultra-high-precision velocity measurements of oscillations in $\alpha$~Cen~A.
From those observations \citet{Bedding04} have derived oscillation amplitudes and also averaged estimates of the oscillation life-times.

The stochastic excitation mechanism has been modeled by several authors \citep[e.g. ][]{GK77,GMK94,Samadi00I}.
On the base of \citet{Samadi00I}'s theoretical model of stochastic excitation  and constraints from a 3D simulation of the Sun,  \citet{Samadi02II}  succeeded in modeling the rates $P$ at which the solar p-modes are excited by using a Lorentzian function as a model for the eddy-time correlation $\chi_k$, instead of using the  Gaussian function.
This result is summarized in Sect.~\ref{sun}.

An open question is then whether or not such non-Gaussian model for $\chi$, is also appropriate for other  solar-like oscillating stars and whether the present model of stochastic excitation is valid for other stars? 
This question is addressed on the basis  of the recent seismic observations of  $\alpha$~Cen~A.

\section{The theoretical model of stochastic excitation}

The model of stochastic excitation we consider in this work  is basically that of \citet{Samadi00I}. This model provides an expression for the rate $P$ at which a given mode with frequency  $\nu_0$ is excited. This expression can be written in a  schematic form as: 
\eqn{
P(\nu_0) \propto \int d{\rm m} \, \int d{\rm r} \, d\tau \, \, \vec \xi \, .  \,\langle \vec S \, \vec S \rangle \, .  \,\vec \xi
}

The calculation of the rate $P(\nu_0)$  at which a given p-mode of frequency $\nu_0$ is excited, then results  from an integration over the stellar mass $m$ and local integrations over  distance $r$ and time $\tau(r)$ of the mode eigenfunction $\xi$ and the correlation product of the excitation sources $\langle S S\rangle$.

The excitation sources, $S$, are : the turbulent Reynolds stress and the advection of the turbulent entropy fluctuations by the turbulent motions.
$<S S>$  is expressed in term of the turbulent kinetic energy spectrum $E(k,\nu)$ where $k$ is the wavenumber of a given turbulent element \citep[see][]{Samadi00I}.
Following \citet{Stein67}, $E(k,\nu)$ is split into a spatial component, $E(k)$, and a frequency component ---  $\chi_k (\nu)$, also designed as the eddy-time correlation function --- so that
\eqn{
E( k,\nu) =E(  k) \, \chi_k(\nu) \; .
\label{eqn:ek_chik}
}

\section{Computation of the excitation rates $P$}
%
%
 We consider a  3D simulation of $\alpha$~Cen~A computed with \citet{Stein98}'s 3D numerical code which models only the upper part of  the convective zones.  The grid of mesh points of the simulation is 50 x 50 x 82 large.  The simulation has a solar metalicity while the real star has a metallicity [Fe/H]~$\sim$~0.2.

We compute  the rate  $P(\nu)$ at which p modes are excited in  $\alpha$~Cen~A according to Eq.~(1) in the manner of \citet{Samadi02I}:
  The 1D stellar model consistent with the simulations of stars is computed with \citet{JCD83a}'s stellar code and assumes the classical mixing-length formulation of convection.  The mixing-length parameter of the 1D model is adjusted in order that the 1D model matches the simulation.
 The eigenfunctions ($\xi$) and their associated frequencies ($\nu$) are computed with \citet{JCD91b}'s adiabatic code.
 The total kinetic energy contained in  the turbulent kinetic spectrum, $E(k)$,  its depth dependence, and its  $k$-dependency are obtained directly from the 3D simulation. 
  For the eddy time-correlation function $\chi_k$, a  Gaussian  function is usually assumed, {\it i.e.}  it is assumed that two distant points in the turbulent medium are un-correlated  \citep[e.g.][]{GK77,Balmforth92c}.  Here, we investigate both a Gaussian and a Lorentzian form for $\chi_k$.

%
%
For the  Sun (see Fig.~1):  the rate  $P(\nu)$ at which solar p modes are excited $P(\nu)$, was already computed in \citet{Samadi02I,Samadi02II} on the basis of a 3D simulation of the Sun with a higher spatial resolution (253x253x163) than that of the present simulation of $\alpha$\,Cen\,A.

\section{Excitation of solar p~modes}
\label{sun}

It is shown --- on the base of a 3D simulation of the Sun  --- that the Gaussian function usually used for modeling the eddy-time correlation function, $\chi_k$, is inadequate \citep{Samadi02II} .  Furthermore the use of the Gaussian form under-estimates  $P$ as shown in Fig.~1 with  the blue curve. 

On the other hand, a  Lorentzian form   fits best the  frequency dependence of $\nu$  as  inferred from   a  3D simulation of the Sun.  Computed values  of $P$ based on the present model of stochastic excitation and on a Lorentzian function for $\chi_k$  reproduce better the solar seismic observations by \citet{Chaplin98} as shown in Fig.~1 with the red curve.

This result then shows that,  provided that such non-Gaussian model is assumed, the   model of stochastic excitation  is --- for the Sun --- rather satisfactory without adjustment of free parameters in contrast with previous approaches \citep{JCD82,Balmforth92c,GMK94,Samadi00II}

\fig{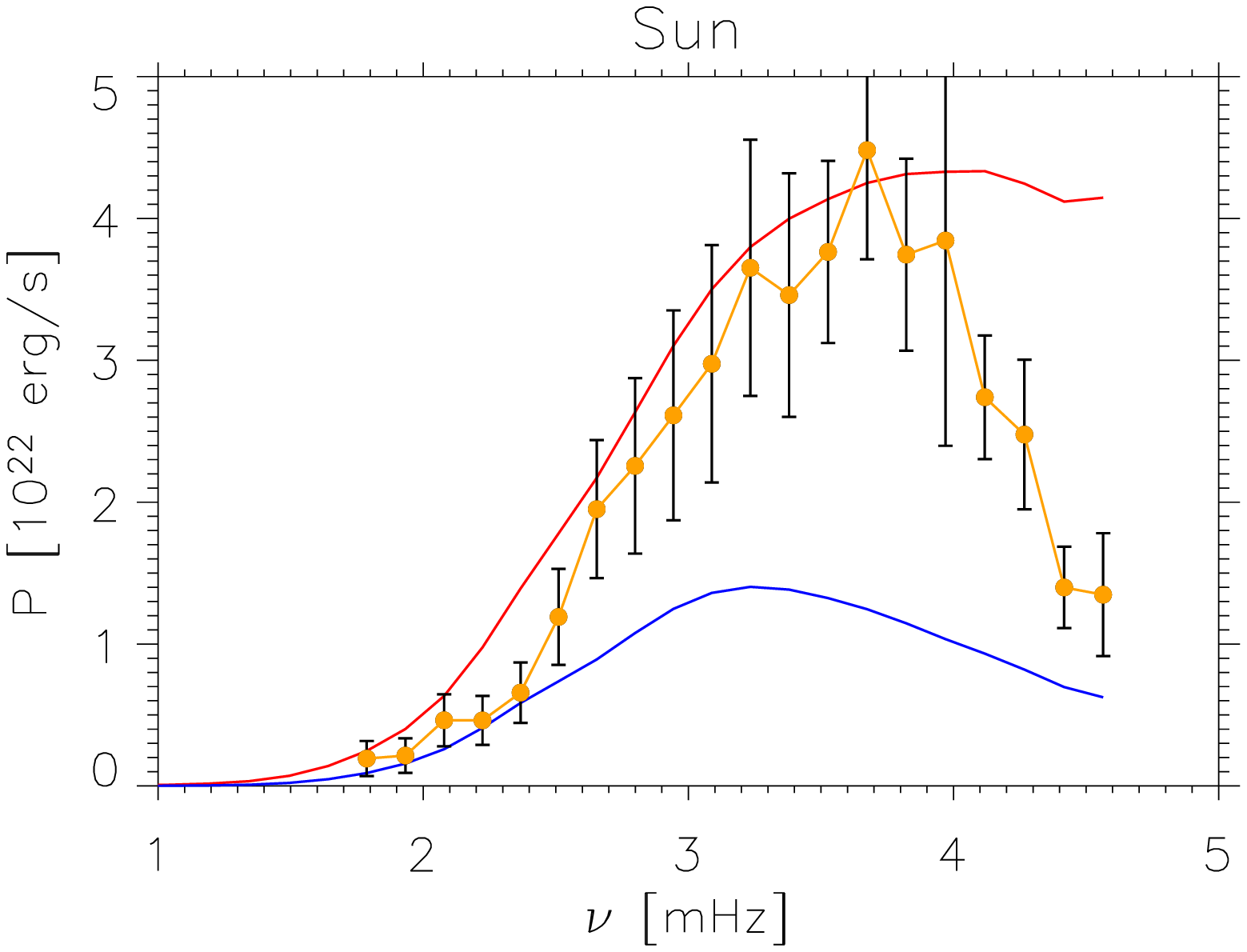}{Rates $P$ at which solar modes with frequency n are excited.
orange: excitation rates $P$ deduced from the seismic observations from the BiSon network \citep{Chaplin98}
blue and red: excitation rates computed as explained in Sects.~2 and 3. The blue (red resp.) curve corresponds to a calculation in which $\chi_k$  is assumed Gaussian (Lorentzian resp.~).}{fig1}

\section{$\alpha$~Cen~A: the observations by Bedding et al (2004)}

Solar-like oscillations have recently been detected in $\alpha$\,Cen\,A by \citet{Butler04} with the UCLES spectrograph. The identification of the modes and the determination of the mode amplitudes have been recently performed by \citet{Bedding04}. Fig. 2 presents the mode amplitudes in Doppler velocity for the l=0,1,2,3 modes which have been detected. 

The authors have also derived --- on the basis of the method proposed by \citet{Stello04} --- an  estimation of the  mode life-times for two  different frequency ranges (see Table 1).

\begin{table}[b]
\caption{Life-time of the modes as derived by \citet{Bedding04}  for two different frequency ranges.}
\label{tab1}
\begin{center}
\begin{tabular}{cccc}  
\hline
Frequency range ($\mu$Hz) & 1700-2400 & 2400-3000     \cr
Mode life-time (days)    & 1.4 & 1.3      \cr
\hline 
\end{tabular}
\end{center}
\end{table} 

\fig{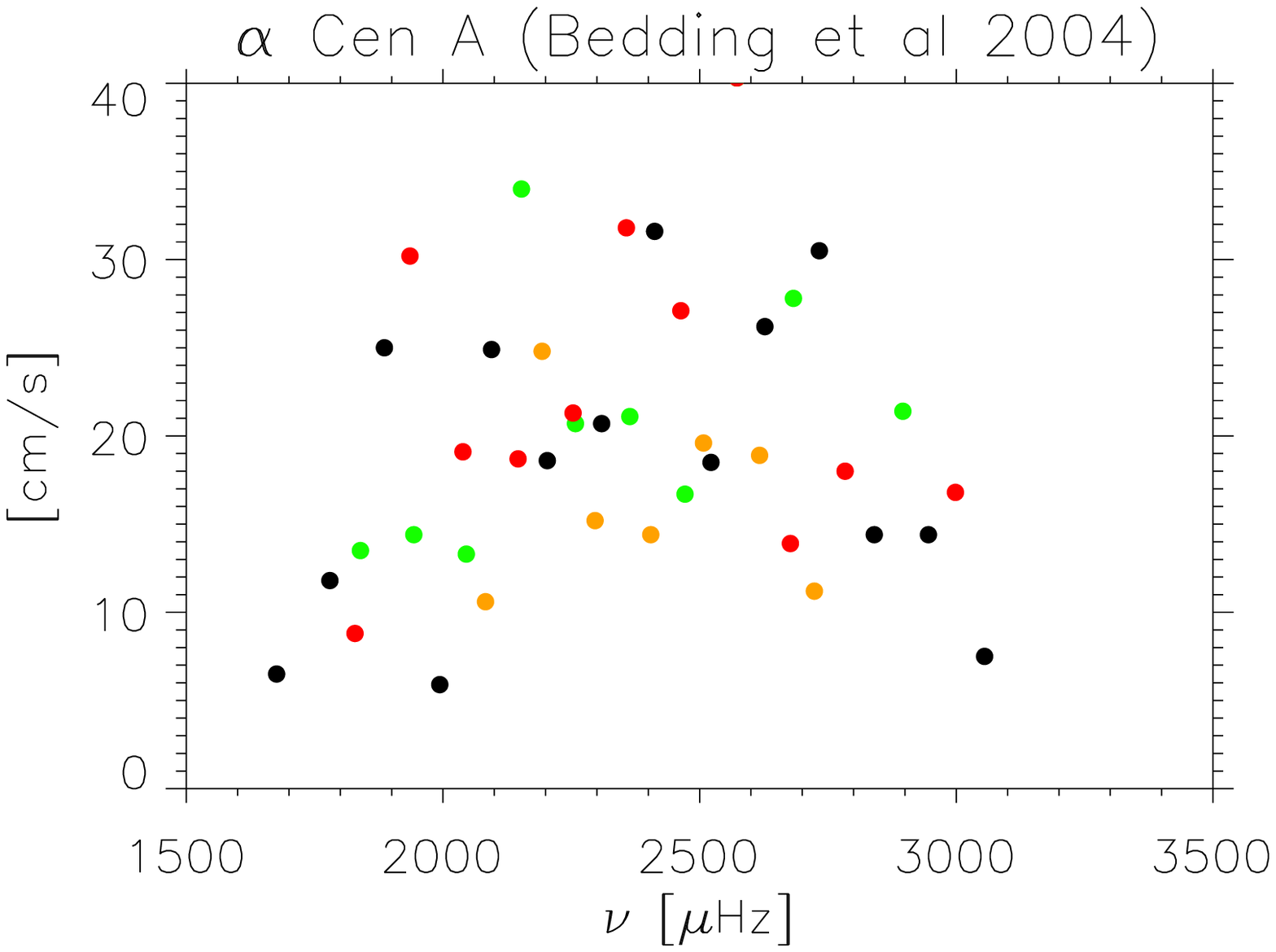}{Amplitudes in Doppler velocity of the modes detected in $\alpha$~Cen~A by \citet{Bedding04}.
The green filled dots correspond to the l=0 modes, the black ones to the l=1 modes, the red ones to the l=2 modes and the orange one to the l=3 modes. Here the observed amplitudes have not been corrected for the mode visibility.}{fig2}

\section{$\alpha$~Cen~A: inferring the intrinsic amplitudes of the modes from observations}

As can been seen in Fig.~2, the measured amplitudes are very scattered. Due to visibility effects, intrinsic amplitudes of l=3 modes ($I3$) are expected to be much smaller than those of l=1  ($I1$)  {\it  i.e.}  $I3 \ll I1$.  

However some observed amplitudes of l=3   ($O3$)   are found as large as those of l=1 ($O1$) but for a given radial order, amplitudes for l=1, l=3 modes are clearly anticorrelated as shown in Fig.~2. This suggests that  the observed  amplitude of a given mode l=3 can be  strongly biaised by the presence of a neighborhood l=1 mode. In contrast, the pollution of the l=1 amplitude  by that of a l=3 one is expected to be quite small. Neglecting this pollution effect, we sum the amplitudes of l=3 and l=1  neighboring modes of a given radial order and attribute this amplitude to l=1 mode {\it i.e.}: 
Since $I3 \ll I1$, then $O3 + O1 = I3 + I1 \sim I1$.  We next correct for the mode visibility.  

We perform the same with the l=0 modes: we sum the   amplitude of l=0 mode and the l=2 mode. However this determination of the amplitudes of the l=0 modes  must be more biased than for the l=1 modes as the amplitudes  of the l=2 modes are not so small compared with those of the l=0 modes.

Results of this determination of the intrinsic mode amplitudes are presented in Fig.~3.

\fig{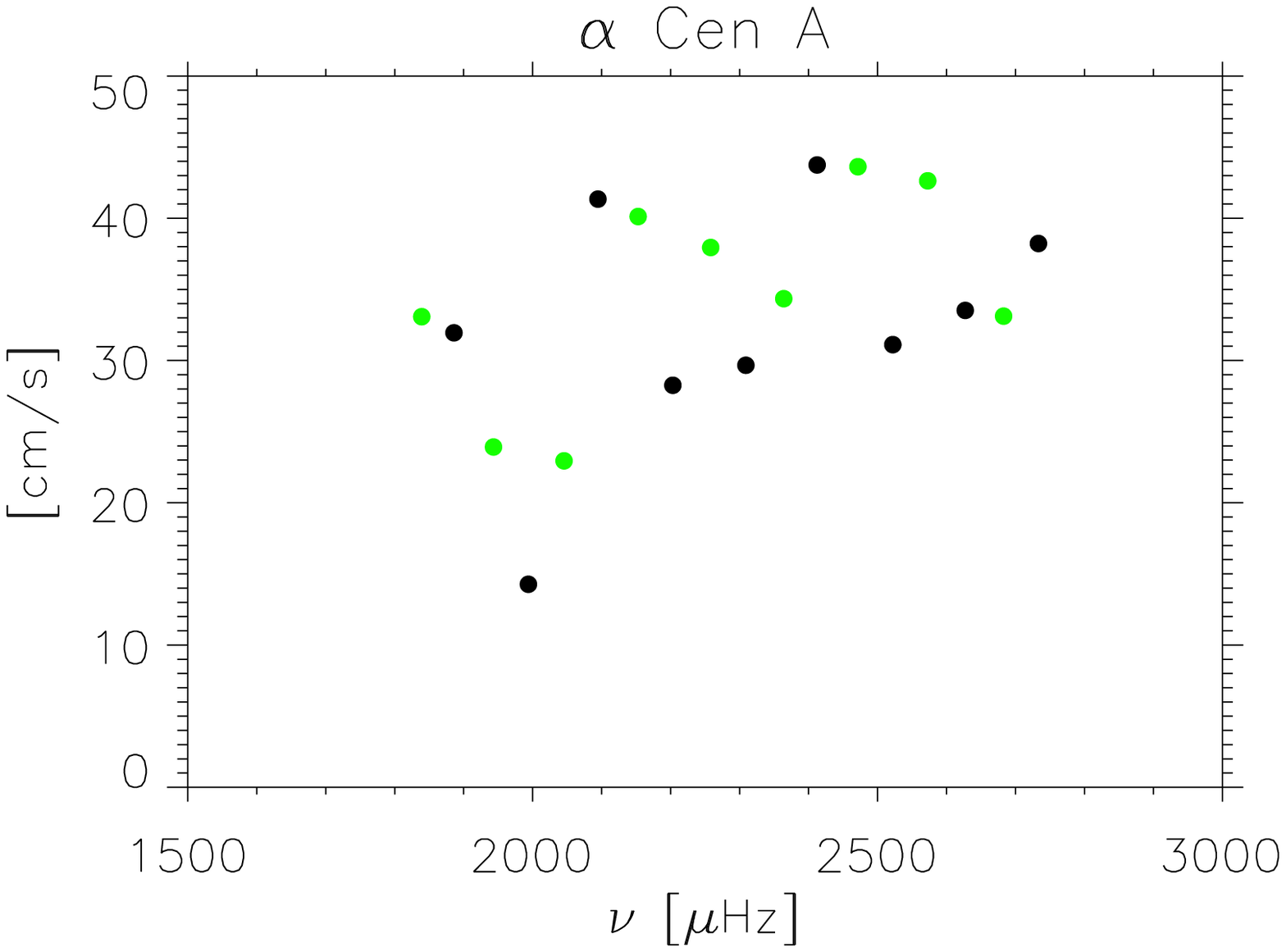}{Inferred intrinsic amplitudes of the l=1 modes (black) and of the l=0 modes (green). The amplitudes of the l=1 modes  are corrected for the mode visibility}{}

\section{$\alpha$~Cen~A:  comparison of the predicted excitation rates $P$ with those inferred from the observations}
We derive the 'observed' excitation rates $P$ from the intrinsic mode amplitudes (Fig.~3) and the mode life-times (Table~1) inferred from the observations according to the relation:
\eqn{
P= 2 \pi ~ M ~ \Gamma ~ V^2
}
where $\Gamma$  is the mode line-width related to the mode life-time $\tau$ (Table 1) as $\Gamma=1/\pi /\tau$, $M$ is the mode mass calculated from the mode eigenfunctions associated with the 1D stellar model of $\alpha$~Cen~A  and finally $V$ is the intrinsic mode amplitude (Fig.~3).

Theoretical $P$ are calculated according to Eq.~(1) and as explained in Sects.~2 and 3.

Inferred and predicted mode excitation rates of  $\alpha$~Cen~A are presented in Fig.~4.

\fig{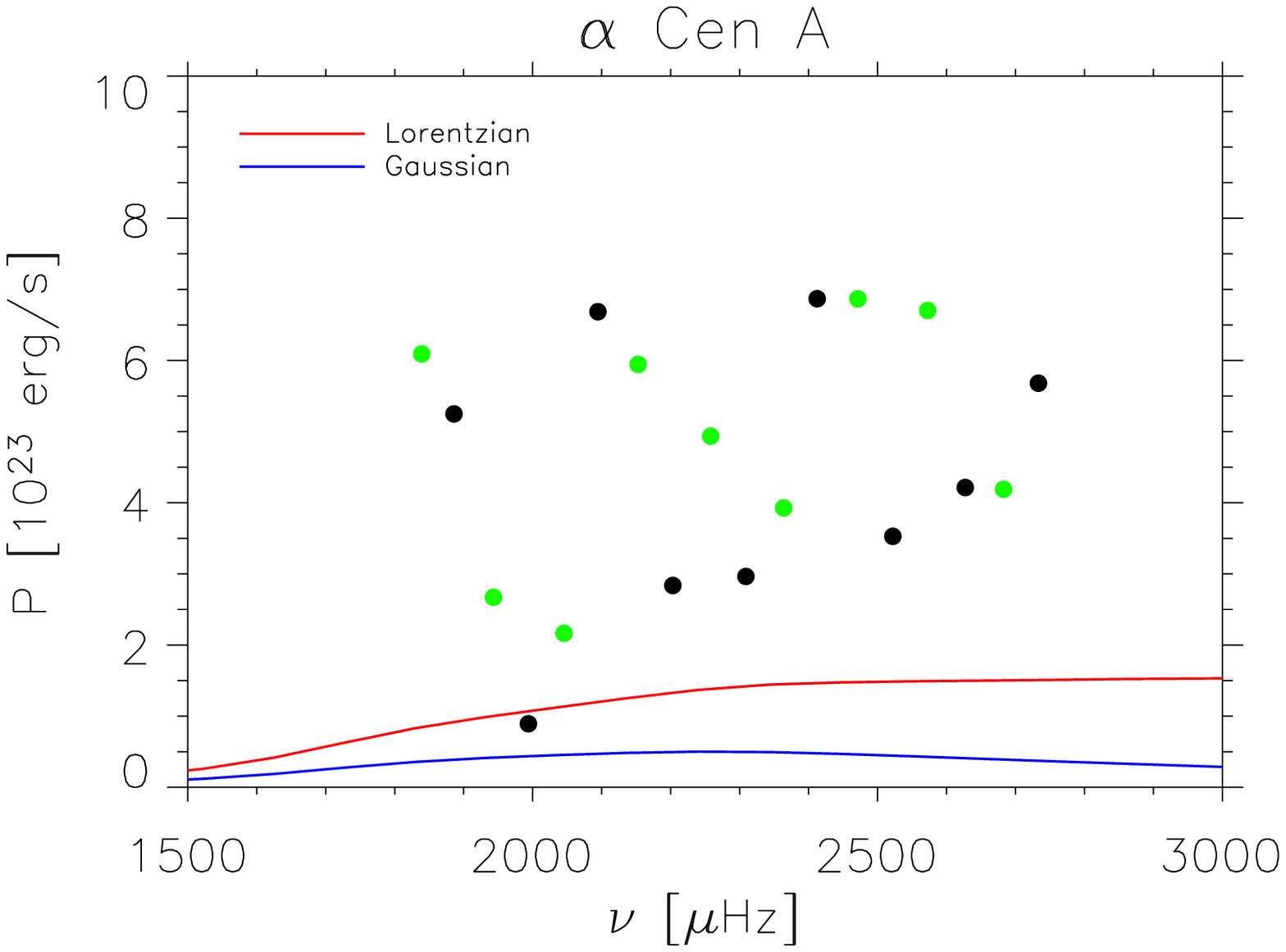}{Filled dots correspond to  the mode excitation rates derived from the inferred intrinsic  amplitude (Fig.~3) of the l=0 mode (green) and the l=1 mode (black)  amplitudes and the estimated mode life-times (Table 1). 
The blue and red curves correspond to the excitation rates computed as explained in Sects. 2 and 3. The blue (red resp.) curve corresponds to a calculation in which $\chi_k$   is assumed Gaussian (Lorentzian resp.). }{}

\section{Conclusion}

For $\alpha$\,Cen\,A as for the Sun,  $P$ computed assuming the Lorentzian function for the eddy-time correlation function   $\chi_k$    tend to be closer to the observations than those assuming the Gaussian function, however the quality of the data does not allow to conclude firmly about the best function  for $\chi_k$.

In this first tentative comparison, the predictions and the observations  of $\alpha$~Cen~A  disagree by a large amount  (the  inferred $P$ are very scattered: the disagreement  varies by a factor 2 to 7 approximatively).
This large discrepancy can either be caused by the present data quality which is not sufficient for our purpose , by the way the intrinsic amplitudes and the life-times of the modes are determined or finally attributed to the present modelling. Data of higher quality or/and  more adapted data reductions are really welcome for deriving precise constraints on the p-mode excitation in $\alpha$~Cen~A.

\section*{Acknowledgments}

 RS acknowledges support by Comit\'e National Francais d'Astronomie and by the Scientific Council of Observatory of Paris.
We thank T. Bedding for having providing us their seismic analysis of   $\alpha$~Cen~A.


\bibliography{../../biblio}
\bibliographystyle{aa}

\end{document}